\newcommand{\version}{November 3, 2011}
\documentclass[11pt,a4paper,english,pdftex]{article}

\usepackage[bindingoffset=0.3cm,textheight=22.5cm,hdivide={2.7cm,*,2.7cm}, vdivide={*,22cm,*}]{geometry}
\usepackage[pdftex]{graphicx}
\usepackage[pdftex,bookmarksnumbered=true,breaklinks=true]{hyperref}

\usepackage{amsmath,amsfonts,amssymb,slashed,mathrsfs}

\usepackage[numbers,square,comma,sort&compress]{natbib}

\IfFileExists{dsfont.sty}
	{\usepackage{dsfont}
         \newcommand{\id}{\mathds{1}}}
	{\typeout{Package dsfont.sty was not found, using alternative macros.}
         \let\mathds=\mathbb
         \newcommand{\id}{\mbox{1 \kern-.59em {\rm l}}}}
\let\one=\id

\newcommand{\nocontentsline}[3]{}
\newcommand{\tocless}[3]{\bgroup\let\addcontentsline=\nocontentsline#1{#2}#3\egroup}

\newcommand{\qed}{\nobreak \ifvmode \relax \else
      \ifdim\lastskip<1.5em \hskip-\lastskip
      \hskip1.5em plus0em minus0.5em \fi \nobreak
      \vrule height0.75em width0.5em depth0.25em\fi}

\newcommand{\be}{\begin{equation}}
\newcommand{\ee}{\end{equation}}
\newcommand{\eq}[1]{(\ref{#1})}

\def\nn{\nonumber}
\def\bea{\begin{eqnarray}}
\def\eea{\end{eqnarray}}
\def\obar{\overline}

\def\beqa{\begin{eqnarray}} 
\def\eeqa{\end{eqnarray}} 
\def\beq{\begin{equation}} 
\def\eeq{\end{equation}}

\def\Tr{{\rm Tr}}

\def\a{\alpha}          
\def\b{\beta}           

\def\l{\lambda} \def\L{\Lambda} 

\def\r{\rho}

\def\cA{{\cal A}}  \def\cC{{\cal C}}
  \def\cF{{\cal F}}
 \def\cH{{\cal H}} 
\def\cJ{{\cal J}} \def\cK{{\cal K}} 
\def\cM{{\cal M}} \def\cN{{\cal N}} \def\cO{{\cal O}}
\def\cP{{\cal P}}

\def\algA{\mathscr{A}}

\newcommand{\sD}{\slashed{D}}

\newcommand{\bc}{\bar c}

\newcommand{\R}{\mathds{R}}

 \newcommand{\msu}{\mathfrak{s}\mathfrak{u}}

\def\bit{\begin{itemize}}
\def\eit{\end{itemize}}

\def\({\left(}
\def\){\right)}

\def\pa{\partial}

\newcommand{\tr}{\mbox{tr}}

\def\bcomment#1{}

\def\LNC{\L_{\rm NC}}
\newcommand{\bLNC}{\bar\L_{\rm NC}}

\def\Pfaff{\rm Pfaff}

\newcommand{\Sin}[1]{\sin\!\left(\! \frac{ #1 }{2} \!\right)}
\newcommand{\Cos}[1]{\cos\!\left(\! \frac{ #1 }{2} \!\right)}

\newcommand{\moyal}{Groenewold-Moyal}
\newcommand{\uim}{UV/IR mixing}
\newcommand{\nc}{non-com\-mu\-ta\-tive}

\newcommand{\eqnref}[1]{Eqn.~(\ref{#1})}		
\newcommand{\figref}[1]{Fig.~\ref{#1}}			
\newcommand{\secref}[1]{Section~\ref{#1}}		
\newcommand{\inv}[1]{\frac{1}{#1}}				

\newcommand{\ri}{{\rm i}}
\newcommand{\pb}[2]{\{#1,#2\}}						
\newcommand{\co}[2]{[#1,#2]}						
\newcommand{\intx}{\int\! d^4x}						
	
\renewcommand{\a}{\alpha}
\renewcommand{\b}{\beta}
\newcommand{\g}{\gamma}
\renewcommand{\d}{\delta}
\newcommand{\e}{\epsilon}

\renewcommand{\th}{\theta}

\renewcommand{\l}{\lambda}
\newcommand{\m}{\mu}
\newcommand{\n}{\nu}

\renewcommand{\r}{\rho}
\newcommand{\s}{\sigma}

\newcommand{\vph}{\varphi}

\newcommand{\G}{\Gamma}

\newcommand{\Th}{\Theta}

\renewcommand{\L}{\Lambda}

\renewcommand{\Xi}{\Xi}

\title{\begin{flushright}
       \end{flushright}
Heat kernel expansion and induced action for matrix models}
\author{Daniel N. Blaschke\footnote{daniel.blaschke@univie.ac.at}}

\date{\version}
\graphicspath{{./figures/}}
\begin{document}

\maketitle

\begin{center}
University of Vienna, Faculty of Physics\\
Boltzmanngasse 5, A-1090 Vienna (Austria)
\vspace{0.5cm}
\end{center}%

\begin{abstract}
In this proceeding note\footnote{based on a talk given at the 7$^{\rm th}$ International Conference on Quantum Theory and Symmetries, August 7--13, 2011 in Prague, Czech Republic}, 
I review some recent results concerning the quantum effective action of certain matrix models, i.e. the supersymmetric IKKT model, in the context of emergent gravity. 
The absence of pathological {\uim} is discussed, as well as dynamical SUSY breaking and some relations with string theory and supergravity. 
\end{abstract}

\section{Background}

One of the greatest challenges of present day physics is to find a mathematical framework to describe quantum field theory formulated on some quantum space-time. 
This would be necessary in order to combine the standard model of particle physics with gravity. 
Surely, many ideas are currently around and it remains unclear which is the best way to achieve this goal. 
What I would like to present in this short proceeding note, are some recent results~\cite{Blaschke:2010rr,Blaschke:2011qu} which are based on the idea of gravity as an emergent force derived from matrix models --- for a general introduction to the subject, the interested reader is referred to~\cite{Steinacker:2010rh}.

It is indeed not easy to find a new model that is renormalizable, but a concept that is known to greatly improve the properties of a model with respect to divergences and renormalizability is supersymmetry.
Incidentally, a supersymmetric matrix model which turns out to achieve what we want, namely incorporating scalars, fermions, gauge bosons and (emergent) gravity was originally introduced as a non-perturbative description of type IIB string theory:
It is the so-called IKKT model, named after its authors~\cite{Ishibashi:1996xs}, and given by the action
\begin{align}
S_{\rm IKKT}&=-(2\pi)^2\Tr\left(\co{X^a}{X^b}\co{X_a}{X_b}\,\, + \, 2\obar\Psi \gamma_a[X^a,\Psi] \right) 
\,,
\label{eq:IKKT-MM}
\end{align}
where $X^a,\,\, a= 0,1,2,\ldots,9$ are Hermitian matrices,
$\Psi$ is a matrix-valued Majorana-Weyl spinor of $SO(9,1)$,
and the $\g_a$ form the corresponding Clifford algebra.
The model is obtained by dimensional reduction of the 10-dimensional $SU(N)$ Super-Yang-Mills theory
to a point, and taking the $N \to\infty$ limit. 
Indices are raised and lowered using the fixed background metric $g_{ab}$, which we consider to either be the ten-dimensional Minkowski or Euclidean metric.
The action \eqref{eq:IKKT-MM} is invariant under the $\cN=2$ matrix supersymmetry
\begin{align}
 \d_{\e}^{(1)}\psi&=\frac i2[X_a,X_b]\G^{ab}\e\,, &
\d_{\e}^{(1)}X_a&=i\bar\e\G_a\psi\,, \nn\\
\d_{\xi}^{(2)}\psi&=\xi\,, &
\d_{\xi}^{(2)}X_a&=0 
\,, \label{eq:IKKT-susy}
\end{align} 
as well as the additional symmetries
\begin{align}
 \begin{array}{lllll}
X^a \to U^{-1} X^a U\,,\quad  &\Psi \to U^{-1} \Psi U\,,\quad  &  U \in U(\cH)\,\quad  & \mbox{gauge invariance,}  \\
X^a \to \L(g)^a_b X^b\,,\quad  &\Psi_\a \to \tilde \pi(g)_\a^\b \Psi_\b\,,\quad  & g \in \widetilde{SO}(9,1)\,\; 
  & \mbox{Lorentz symmetry,}  \\
X^a \to X^a + c^a \one\,,\quad & \quad & c^a \in \R\,\quad  & \mbox{translational symmetry,}
\end{array}
\label{transl-inv}
\end{align}
where the tilde indicates the corresponding spin group.
Although an explicit proof to all orders is still missing, there are good reasons to believe that this model is in fact renormalizable~\cite{Ishibashi:1996xs,Jack:2001cr}. 

Here we take the IKKT matrix model model \eqref{eq:IKKT-MM} as a starting point independent of string theory, and focus on 4-dimensional brane solutions or backgrounds, considered as physical space(-time). 
Due to  supersymmetry, the model should provide a well-defined quantum theory at least for 4-dimensional backgrounds. 
In fact, numerical evidence for the emergence of 3+1-dimensional space-time within the IKKT model has been obtained recently in Ref.~\cite{Kim:2011cr}, providing further motivation to study the effective physics of 4-dimensional backgrounds. 

Additionally, it was shown in~\cite{Chatzistavrakidis:2011gs} how the particle spectrum of the standard model may be correctly reproduced from that action by considering specific brane solutions, hence showing how realistic physics may emerge from the IKKT model. 

The main emphasize of this article, however, will be on the one-loop effective action of the IKKT model.

\section{The idea of emergent gravity from matrix models}

\begin{figure}[ht]
\centering
\includegraphics[scale=1.0]{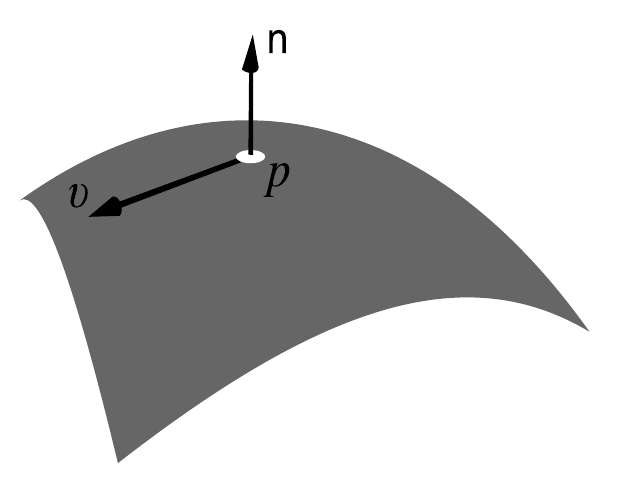}
{\setlength{\unitlength}{1pt}
\put(-145,40){$g_{\mu\nu}(x)$}
\put(-70,40){$\cM^{2n}$}
}
\caption{Embedding and induced metric; projectors on the tangential/normal bundle of $\cM^4\in\R^D$ may be defined as 
$\cP^{ab}_T=g^{\m\n}\pa_\m x^a\pa_\n x^b $ and $\cP^{ab}_N=\eta^{ab}-\cP^{ab}_T$.}
\label{fig:embedding}
\end{figure}

Before we proceed, we ought to explain how gravity may emerge from such matrix models (cf.~\cite{Steinacker:2010rh,Blaschke:2010rg,Blaschke:2010qj}).
This is best shown by considering the bosonic part of the action \eqref{eq:IKKT-MM}, namely
\begin{align}
S_{YM}&=-\Tr\co{X^a}{X^b}\co{X^c}{X^d}\eta_{ac}\eta_{bd}
\,. \label{YM--model}
\end{align}
The $X^a$ are Hermitian matrices on $\mathcal{H}$ which in the semi-classical limit are interpreted as coordinate functions. If (in the simplest case) one considers some of the coordinates to be functions of the remaining ones~\cite{Steinacker:2008ri}
such that $X^a \sim x^a = (x^\mu,\phi^i(x^\mu))$ in the semi-classical limit,
one can interpret the $x^a$ as defining the embedding of a $2n$-dimensional submanifold 
$\cM^{2n}\hookrightarrow\R^D$ equipped with a non-trivial induced metric (cf. \figref{fig:embedding})
\begin{align}
g_{\m\n}(x)&=\pa_\m x^a \pa_\n x^b\eta_{ab}
=\eta_{\m\n}+\pa_i\phi^i(x)\pa_j\phi^j(x)
\,, 
\end{align}
via pull-back of $\eta_{ab}$, and where $\m,\n\in{1,\ldots,2n}$ and $i,j\in{2n+1,\ldots,D}$. 
Here we consider this submanifold to be a four dimensional space-time $\cM^4$.
However, it is not the induced metric which is ``seen'' by scalar fields, gauge fields, etc., but the effective metric~\cite{Steinacker:2007dq}
\begin{align}
G^{\m\n}&=e^{-\s}\th^{\m\r}\th^{\n\s}g_{\r\s}
\,, &
e^{-\s}&\equiv \frac{\sqrt{\det\th^{-1}_{\m\n}}}{\sqrt{\det G_{\r\s}}}\,.
\end{align}
This can be best understood from the following simple example: The gauge invariant kinetic term of a test particle modelled by a scalar field $\phi$ has the form
\begin{align}
S[\phi]&=-\Tr\co{X^a}{\Phi}\co{X^b}{\Phi}\eta_{ab} \nn\\
&\sim\intx\sqrt{\det\th^{-1}}\th^{\m\n}\pa_\m x^a\pa_\n\phi\th^{\r\s}\pa_\r x^b\pa_\s\phi\eta_{ab} \nn\\
&=\intx\sqrt{\det G}G^{\n\s}\pa_\n\phi\pa_\s\phi
\,.
\end{align}
The semi-classical expression on the r.h.s. clearly describes a scalar particle in a curved background with metric $G_{\m\n}$, namely the effective metric defined above.

Furthermore, we can interpret
\begin{align}
-\ri\co{X^\m}{X^\n}\sim \pb{x^\m}{x^\n}= \ri\th^{\m\n}(x)
\end{align}
as a Poisson structure on $\cM^4$. We assume that $\th^{\m\n}$ is non-degenerate, 
so that its inverse matrix $\th^{-1}_{\m\n}$ defines a symplectic form 
\begin{align}
\varTheta=\inv{2}\theta^{-1}_{\mu\nu} dx^\mu \wedge dx^\nu
\end{align}
on $\cM^4$. 
In fact, the special case of a (anti-) self-dual symplectic form, i.e. $\star\varTheta=\pm\ri\varTheta$, leads to the interesting relation $G_{\m\n}=g_{\m\n}$. 
This means, that the induced and the effective metric coincide. 
Note, however, that this is true only for 4-dimensional embedding spaces $\cM^4$ since in that case the determinants of the metrics already coincide, i.e. $|G|=|g|$.
(For details, see~\cite{Steinacker:2010rh}.) 
Moreover, this special case corresponds to a K\"ahler manifold, as can be seen from the following considerations:
We introduce the new quantity $\cJ^\m_{\ \n}$ by writing the effective metric as
\begin{align}
 G^{\m\n}=:-(\cJ^2)^\m_{\ \r}g^{\r\n}
 \,, \label{eq:def-semiJ}
\end{align}
and hence 
\begin{align}
 \cJ^\m_{\ \n}&=e^{-\s/2}\th^{\m\r}g_{\r\n}\,, &
 (\cJ^2)^\m_{\ \r}&=-G^{\m\s}g_{\s\r}
 \,. 
\end{align}
Clearly, one has $\cJ=-\id$ if $G^{\m\n}=g^{\m\n}$ meaning it is an almost complex structure.
In general, $\cJ$ fulfills the following characteristic equation (for 4-dimensional $\cM^4$):
\begin{align}
 {(\mathcal{J}^2)^\mu}_\nu + \frac{(Gg)}{2} {\delta^\mu}_\nu + {(\mathcal{J}^{-2})^\mu}_\nu = 0
\,. \label{eq:characteristic-J}
\end{align}
In fact, this relation will become important in deriving \eqnref{eq:matrix-eff-action_fermionic} in \secref{sec:fermion-abelian-induced}.

\section{Deriving the effective action}

\paragraph{The fermionic part.} 
As we have already mentioned, the fermions in the IKKT model, which is formulated on $9+1$ dimensional Minkowski space-time, are Majorana-Weyl spinors $\Psi_C=\cC\bar\Psi^T=\Psi$, i.e. the spinor entries are Hermitian matrices.
Hence, the fermionic action can be written as
\begin{align}
\Tr\bar\Psi\g_a\co{X^a}{\Psi}&=\Tr\Psi\cC\g_a\co{X^a}{\Psi}
\,, 
\end{align}
and
\begin{align}
e^{\ri\G^\psi[X]}:=\int\!d\Psi e^{\ri\Tr\bar\Psi\g_a\co{X^a}{\Psi}} 
=\Pfaff(\cC\g_a\co{X^a}{.}=\pm\sqrt{\det(\cC\sD_+)}
\,, 
\end{align}
where $\sD$ denotes the Dirac operator acting on the positive chirality spinors. 
In fact, this expression may be used to define the Wick-rotated Euclidean fermionic induced action $\G^\psi_E[X]$. One replaces $\g_0\to\ri\g_{10}$ leading to a non-Hermitian operator $\cC\sD$. 
Hence, the effective Euclidean action has both a real and an imaginary contribution. 
The latter is the Wess-Zumino contribution~\cite{Belyaev:2011dg} which we will not discuss here. 
The real part, on the other hand, can be written as
\begin{align}
\G^{\psi,\textrm{real}}_E[X]&=-\inv{4}\Tr\log(\sD^2)
=\inv{4}\Tr\int_0^\infty\frac{d\a}{\a}e^{\a(\Box+\Sigma^{(\psi)}_{ab}\co{\Th^{ab}}{.}}
\,, 
\end{align}
where 
\begin{align}
\Box&=\co{X^a}{\co{X_a}{.}}\,, &
\Th^{ab}&=-\ri\co{X^a}{X^b}\,, \nn\\ 
(\Sigma^{(\psi)}_{ab})^\a_\b&=\frac{\ri}{4}\co{\g_a}{\g_b}^\a_\b
\,. 
\end{align}
For simplicity, we consider the Euclidean case in the following.

\paragraph{The bosonic part.}
In order to compute the effective action of\footnote{We remain in Euclidean space.}
\begin{align}
S_{\textrm{YM}}&=-(2\pi)^2\Tr\left(\co{X^a}{X^b}\co{X_a}{X_b}\right)
\,, \label{eq:bosonic-action-inv}
\end{align}
we employ the background field method~\cite{Peskin:1995,Weinberg:1980wa,Khoze:2000sy} and hence split the matrices as $X^a\to X^a+Y^a$, where $Y^a$ denotes the fluctuating part. 
The gauge symmetry $Y^a\to Y^a+U\co{X^a+Y^a}{U^{-1}}$ needs to be fixed using a gauge fixing function, which we take to be $G[Y]=\ri\co{X^a}{Y_a}$. 
Together with the according ghost-part, the additional terms read
\begin{align}
S_{\textrm{gf}+\textrm{ghost}}&=-2(2\pi)^2\Tr\left(\co{X^a}{Y_a}\co{X^b}{X_b}-2\bc\co{X^a}{\co{X_a+Y_a}{c}}\right)
\,. \label{eq:MM-gf}
\end{align}
For the one-loop computation of the effective action for $X^a$ one keeps only the quadratic parts in the fluctuations $Y^a$. For the sum of \eqref{eq:bosonic-action-inv} and \eqref{eq:MM-gf} this leads to
\begin{align}
S_{\text{quad}}&=2(2\pi)^2\Tr\left(Y^a(\Box+\Sigma^{(Y)}_{cd}\co{\Th^{cd}}{.})^a_bY^b+2\bc\Box c\right)
\,, 
\end{align}
where 
\begin{align}
(\Sigma^{(Y)}_{cd})^a_b&=\ri(\d^a_cg_{bd}-\d^a_dg_{bc})
\,. 
\end{align}
The effective action due to the bosonic fluctuations $Y$ hence becomes
\begin{align}
\Gamma^{Y+c}[X]&=-\inv{2}\Tr\log(\Box+\Sigma^{(Y)}_{cd}\co{\Th^{cd}}{.})+\tr\log\Box
\,. 
\end{align}

Putting everything together, we are now ready to write down the complete (real part of the) one-loop effective action for the IKKT model using a Schwinger parametrization:
\begin{align}
\Gamma[X]=- \frac 12 \Tr \int_0^\infty \frac {d\a}{\a} 
   \Big( e^{-\a(\Box  + \Sigma^{(Y)}_{ab}[\Th^{ab},.])}
  - \frac 12 e^{-\a(\Box  + \Sigma^{(\psi)}_{ab}[\Th^{ab},.]) }
 - 2 e^{-\a\Box}  \Big) 
\,. \label{eq:effective-action-exact}
\end{align}
Due to supersymmetry, a UV cutoff is not required. 
However, once SUSY is broken (e.g. spontaneously) a regularization is needed. 
Hence, one may supplement the $\a$ integral with a ``mild'' cutoff $L$ by multiplying the integrand with $e^{-\inv{\a L^2}} $. 
Having, dimension of length, $L$ essentially sets a lower limit for the $\a$ integral. 
In a background with {\nc} scale $\LNC$ this amounts to a UV cutoff $\L:=\LNC^2L$. 
In \secref{sec:fermion-abelian-induced} we will focus on the fermionic part by itself, and hence make use of this regularization. 

Note, that the effective action \eqref{eq:effective-action-exact} may alternatively be also written as
\begin{align}
\Gamma[X]=\inv2 \Tr \left(\log(\id  + \Sigma^{(Y)}_{ab}\Box^{-1}[\Th^{ab},.])
-\inv2 \log(\id  + \Sigma^{(\psi)}_{ab}\Box^{-1}[\Th^{ab},.])\right) 
\,, \label{eq:effective-action-pert}
\end{align}
which upon expansion of the logs becomes
\begin{align}
\Gamma[X]=\inv2 \Tr \Bigg(\!\! -\inv4 (\Sigma^{(Y)}_{ab} \Box^{-1}[\Th^{ab},.] )^4 
 +\inv8 (\Sigma^{(\psi)}_{ab} \Box^{-1}[\Th^{ab},.])^4 \,\, +  \cO(\Box^{-1}[\Th^{ab},.])^5 \! \Bigg)
\,. 
\end{align}
As a result of the maximal supersymmetry of the model, the $\Sigma^n$-terms with $n<4$ have cancelled in this expansion. 
Furthermore, the model is one-loop finite on 4-dimensional backgrounds and free of pathological {\uim}. 
Note also, that this expression is background independent and applies both to the Abelian and to the non-Abelian case.

\paragraph{Background expansion.}
Consider a background of slowly varying scalar and gauge fields around the {\moyal} space $\R_\th^4$, i.e.
\begin{align}
X^a &= \begin{pmatrix}
 \bar X^\mu \\ 0
\end{pmatrix} \, + \, 
\begin{pmatrix}
 \cA^\mu  \\   \phi^i
\end{pmatrix}
=\begin{pmatrix}
        \bar X^\m-\bar\th^{\m\n}A_\n \\ \LNC^2\vph^i
       \end{pmatrix} \,,\nn\\
\co{\bar X^\m}{\bar X^\n}&=\ri\bar\th^{\m\n}=\textrm{const.} 
\label{eq:general-X}
\end{align}
The effective action \eqref{eq:effective-action-exact} is then computed by integrating out the fermions $\psi$, the ghosts $c,\bc$ and the background fields $Y$, respectively. 
Note, that we have introduced a {\nc} scale $\LNC:=\det\th^{-1}_{\m\n}$ above, which in the present case is a constant. 

The components of the full commutator $\co{X^a}{X^b}$ hence read
\begin{align}
 [X^\mu,X^\nu]&=i(\bar\theta^{\mu\nu}+\mathcal{F}^{\mu\nu})\,,
\qquad\qquad
[X^\mu,\phi^i]=\ri\bar\theta^{\mu\nu}D_\nu\phi^i\,, \nn\\
\mathcal{F}^{\mu\nu}&=-\bar\theta^{\mu\rho}\bar\theta^{\nu\sigma}(\partial_\rho A_\sigma-\partial_\sigma A_\rho-\ri[A_\rho,A_\sigma])\,, \nn\\
D_\nu\phi&=\partial_\nu\phi+\ri[A_\nu,\phi] \,, \nn\\
\co{X^\m}{X^i}&=\th^{\m\n}D_\n\phi^i
\,. 
\end{align}
Furthermore, the generalized Laplacian $\Box$ can be decomposed as
\begin{align}
\Box\Psi&=\co{X^a}{\co{X_a}{\Psi}}=\bar\Box\Psi+V\Psi
\,, 
\end{align}
where 
\begin{align}
\bar\Box\Psi&=\eta_{\mu\nu} [\bar X^\mu,[\bar X^\nu,\Psi]] 
  = -\LNC^{-4}\bar{G}^{\mu\nu}\partial_\mu\partial_\nu \Psi 
  \,,
\end{align}
is the free Laplace operator on $\R_\th^4$. 
The explicit computations may be done using a Duhamel expansion around the {\moyal} plane:
\begin{align}
\frac1{2}\Tr\left(\log\Box^2-\log\bar\Box^2\right) &\to-\frac1{2}\Tr\int\limits_0^\infty\frac{d\alpha}{\alpha}
\left(e^{-\alpha\Box^2}-e^{-\alpha\bar\Box^2}\right) e^{-\frac{\LNC^4}{\alpha \Lambda^2}} 
\nn\\
&= \Lambda^4\sum_{n \geq 0} \int d^4 x\, \mathcal{O}\left(\frac{(p,A,\varphi)^n}{(\Lambda,\LNC)^n}\right)
\,, 
\end{align}
where $\L$ denotes an ultraviolet cutoff, which is of course only required once we break supersymmetry. 
This may be achieved by considering a general background geometry $\cM=\cM^4\times\cK$ where $\cK$ denotes some compact manifold. 
A gravitational action is then induced on the brane below the SUSY breaking scale, as suggested in \cite{Steinacker:2008ri}. 
Some simple examples of possible geometries are discussed in Ref.~\cite{Steinacker:2011wb} --- see also~\cite{Chatzistavrakidis:2011su} for related discussions. 

In \secref{sec:fermion-abelian-induced} we will consider only the fermionic action in order to get an idea of what types of terms will be generated without SUSY cancellations, and we will of course need a UV cutoff there. 
In fact, it is what we demand of this cutoff, that differs the present expansion from previous work:
It is known~\cite{Vassilevich:2005vk,Gayral:2006vd}, that taking $\L\to\infty$ while keeping $\LNC$ fixed leads to pathological behaviour of the heat kernel expansion such as problematic {\uim}. 
Hence, we demand~\cite{Blaschke:2010rr} that the quantity 
\begin{align}
 \epsilon(p):=p^2\Lambda^2/\LNC^4\ll1 
 \,, 
\end{align}
be a small parameter, i.e. $\L$ must remain smaller than the {\nc} scale. 
This avoids the problems mentioned above, and is well motivated by the fact, that we need to break the supersymmetry of the IKKT model. 
It makes sense to assume that $\L$, representing the scale of SUSY breaking, is smaller than the {\nc} scale, which in turn would possibly be near the Planck scale. 

Under these assumptions we may expand typical {\uim} terms as
\begin{align}
 e^{-p^2\LNC^4/\alpha}\approx\sum\limits_{m\geq0}a_m\epsilon(p)^m
 \,, 
\end{align}
which hence present no problem at all.
We hence end up with an expansion in three small parameters:
\begin{align}
 \Gamma \sim \Lambda^4\sum_{n,l,k\geq 0} \int d^4 x\, \mathcal{O}\left(\epsilon(p)^n (\frac{p^2}{\LNC^2})^l(\frac{p^2}{\Lambda^2})^k \right)
\,. 
\end{align}
The next section is dedicated to some important results regarding this expansion for the fermionic part of the IKKT action.

\section{Fermion induced action}
\label{sec:fermion-abelian-induced}
For simplicity, we consider the Abelian case in this section, and study the quantum effective action due to fermions. 
We then elucidate the results of Ref.~\cite{Blaschke:2010rr}, that this action can be recast in the form of generalized matrix models\footnote{While the quantization of Yang-Mills matrix models has been studied before e.g. in~\cite{Ishibashi:1996xs,Aoki:1999vr,Imai:2003jb,CastroVillarreal:2005uu,Steinacker:2003sd,Azuma:2005pm}, the results available so far are not very explicit, and not in the form of generalized matrix models.}. 
For this purpose we employ the Weyl quantization map to perform the explicit computations, 
i.e. we map plane waves to generalized eigenfunctions $|p\rangle=e^{\ri p_\m\bar X^\m}\in\cA$ with
\begin{align}
 \bar P_\mu |p\rangle &=\ri p_\mu|p\rangle\,, \qquad \textrm{where} \ \bar P_\mu=-\ri\theta^{-1}_{\mu\nu}[\bar X^\nu,.]\,,  \nn\\
\langle q|p\rangle &= \Tr(|p\rangle\langle q|) = 
 \Tr_\mathcal{H} (e^{-\ri q_\mu \bar X^\mu} e^{\ri p_\mu \bar X^\mu})
= (2\pi\LNC^2)^2   \delta^4(p-q) 
\,. 
\end{align}
The Fourier expanded fields are defined by
\begin{align}
 \Psi=\int\!\!\frac{d^4p}{(2\pi\bLNC^2)}\Psi(p)e^{\ri p_\m \bar X^\m}
 \,. 
\end{align}
Details of the full order-by order computations of the Duhamel expansion may be found in Ref.~\cite{Blaschke:2010rr}; here we only state some particularly interesting results.
For example, adding up the first three order contributions leads to the order $\L^4$ terms
\begin{align}
 \Gamma_{\Lambda^4}(A,\varphi,p) &= \frac{\L^4\Tr\one}{16\LNC^4}\int\!\!\frac{d^4x}{(2\pi)^2} \sqrt{g}
\Big(g^{\alpha\beta}D_\alpha\varphi^i D_\beta\varphi_i 
 - 2\bar\th^{\nu\mu}F_{\m\a} g^{\a\b}\partial_\nu\varphi^i\partial_\b\varphi_i   \nn\\
& \quad - \frac 12\LNC^4 \big( \bar\theta^{\mu\nu} F_{\nu\mu}\bar\theta^{\rho\sigma} F_{\sigma\rho}
 + (\bar\th^{\s\s'}F_{\s\s'}) (F\bar\th F\bar\th)\big)
 +\frac 12(\bar\th^{\mu\nu}F_{\mu\nu}) g^{\a\b}  \partial_\b\varphi^i\partial_\a\varphi_i  
 \nn\\ &\quad
 + \textrm{h.o.}\Big)
\,, \label{eq:eff-L4-3order-fermionic}
\end{align}
which are manifestly gauge invariant.
Remarkably, this result can be precisely recovered from the simple matrix model effective action \eq{eq:matrix-eff-action_fermionic} below. 
%
In fact, the free contribution, which is given by
\begin{align}
\Gamma[\bar X]
&= -\frac1{2}\Tr\int\limits_0^\infty\frac{d\a}{\a} e^{-\a\sD_0^2-\frac{\LNC^4}{\a \L^2}}  
=  -\frac{\L^4 \Tr \one}{8} \int\!\! \frac{d^4 x}{(2\pi)^2}\,\sqrt{g}
\,, \label{eq:free-contrib}
\end{align}
along with general geometrical considerations, suffice to predict such loop computations:
Consider the effective action in terms of matrices. 
Taking into account the scaling property $\Gamma_L[X]=\Tr\mathcal{L}\left(X^a/L\right)$, where $L=\L/\LNC^2$, and that commutators correspond to derivative operators for gauge fields, 
it was shown in Ref.~\cite{Blaschke:2010rr} that the leading term of the effective matrix action can be written in terms of products of
\begin{align}
 J^a_b := i\Theta^{ac}g_{cb} \, = \, [X^a,X_b] , \qquad  \Tr J \equiv J^a_a = 0
 \,. 
\end{align} 
Note, that the semiclassical limit of $J$ has already been introduced in \eqref{eq:def-semiJ} (up to a scale factor $\LNC$), where it was identified as an almost complex structure iff $G_{\m\n}=g_{\m\n}$. 
Furthermore, using the input from the characteristic semi-classical relation \eqref{eq:characteristic-J} and the free contribution to the effective action, it was shown in Ref.~\cite{Blaschke:2010rr} that the most general single-trace form of the effective action is given by 
\begin{align}
&\Gamma_L[X]  = -\frac 14 \Tr \bigg(\frac{L^4}{\sqrt{-\Tr J^4 + \frac 12 (\Tr J^2)^2}}\bigg) 
 + \textrm{h.o.}
 \label{eq:matrix-eff-action_fermionic}
\end{align}
In the semi-classical limit this yields exactly \eqref{eq:free-contrib} for the vacuum. 
When including gauge fields in the computation, \eqref{eq:matrix-eff-action_fermionic} reproduces the result \eqref{eq:eff-L4-3order-fermionic}. 
This implies, that the effective action can be written as a generalized matrix model with manifest $SO(D)$ symmetry, demonstrating the power of the geometric view of the matrix model. 
One should also note at this point, that it is easily possible to further generalize the model to include curvature terms of type
\begin{align}
\int\!\frac{d^4x}{(2\pi)^2}\sqrt{g}\, \L(x)^2 \Big(R  
+ (\bar\L_{\rm NC}^4 e^{-\sigma}\theta^{\mu\rho} \theta^{\eta\alpha} R_{\mu \rho\eta\alpha} - 4 R )
+ c' \partial^{\mu}\sigma \partial_\mu\sigma \Big)
\,, 
\end{align}
which should also appear, as was discussed in~\cite{Steinacker:2008a,Klammer:2009dj,Blaschke:2010qj,Blaschke:2010rr}. 
These would then be higher-order commutator terms, as indicated by ``h.o.'' in \eqnref{eq:matrix-eff-action_fermionic}.

\section{Non-Abelian sector}
Non-Abelian gauge fields appear when one considers a background of $N$ coinciding branes, i.e.
\begin{align}
X^a &= \bar X^a \, \one_N + \cA^a \qquad \in \algA \otimes \msu(N) \,,  \nn\\
\Theta^{ab} &= \bar \Theta^{ab}\one_N  + \cF^{ab}_\a \l^\a 
\,. \label{nonabel-background}
\end{align}
The main difference to the Abelian case concerning loop computations can be seen from the form of a typical vertex term. It reads
\begin{align} 
[\cF^{ab}_\a(k_1) e^{i k_1 X}\l^\a,f_\b(k_2) e^{i k_2 X}\l^\b] 
&= -i\Sin{k_1\th k_2} \cF_\a^{ab}(k_1) f_\b(k_2)e^{i (k_1+k_2) X}  \{\l^\a,\l^\b\} \nn\\
& \qquad +  \Cos{k_1\th k_2}\cF_\a^{ab}(k_1) f_\b(k_2) e^{i (k_1+k_2) X} [\l^\a,\l^\b]  \nn\\
&\stackrel{k_i \to 0}{\rightarrow} 
\cF_\a^{ab}(k_1) f_\b(k_2)e^{i (k_1+k_2) X}  [\l^\a,\l^\b]
\,, 
\end{align}
leading to the same $\msu(N)$-valued contributions as in the commutative case in the low energy limit  $k_i \to 0$, while NC effects are sub-leading. 
This means, that in the supersymmetric case one may expect the low-energy effective action to reduce to $\cN=4$ SYM on a general background $\cM$.

In Ref.~\cite{Blaschke:2011qu}, it was shown that the effective action for an unbroken massless gauge boson in  the Coulomb phase on a generic curved brane leads to the leading terms of the Dirac-Born-Infeld (DBI) action for a single brane at some distance of the stack of $N-1$ branes: 
%
Consider $SU(N)$ broken down to $SU(N-1)\times U(1)$ through scalar fields $\phi^i\sim\lambda$ where $\lambda$ is the generator of the unbroken $U(1)$.
Then the one-loop effective action agrees with expansion of the Dirac-Born-Infeld (DBI) action for a D3-brane  in the background of $N-1$ coinciding branes\footnote{For the detailed computations in this so-called Coulomb branch, we once more refer to Ref.~\cite{Blaschke:2011qu}.}
\begin{align}
 S_{\rm DBI} &=
T_3\! \int_\mathcal{M} \!d^4 x\, \frac{|\phi^2|^2}{Q}\Bigg(\sqrt{\bigg|\det\!\Big(G_{\mu\nu} + \tfrac{Q}{|\phi^2|^2}D_\mu\phi^i D_\nu\phi_i
+ \tfrac{Q^{1/2}}{|\phi^2|} \frac{F_{\mu\nu}}{\LNC^{2}}\Big)\bigg|} -\sqrt{|\det G|}\Bigg), \nn \\
 Q&=\frac{(N-1)}{2\pi^2\LNC^4}\,, 
\qquad\qquad T_3=\LNC^4
\,. \label{DBI-action-AdS}
\end{align}
In string theory, these constants are known to be $Q=\frac{(N-1) g_s{\a'}^2}{\pi}$ and $T_3 = \frac 1{2\pi g_s{\a'}^2}$ \cite{Tseytlin:1999dj}, which is consistent with the above result. 
In fact, it allows to identify the scale of non-commutativity with the string theory parameters as 
\begin{align}
g_s\alpha'^2=\frac1{2\pi\LNC^4}
\,. 
\end{align}
The geometrical meaning of \eq{DBI-action-AdS} may be understood as follows: 
A single D3 brane is modelled by the unbroken $U(1)$ component, whose displacement in the transversal $\R^6$ is given by $\phi^i$. Hence, for small transversal distance $|\phi|$ or large $N$ and $G_{\mu\nu} = \eta_{\mu\nu}$, the  action \eq{DBI-action-AdS} reduces to the DBI action on a geometry with effective metric
\begin{align}
ds^2 &= H^{-1/2}(x) dx^\mu dx_\mu + H^{1/2}(x)(d\phi^2 + \phi^2 d\Omega^5)\,, \nn\\
H &= 1+ \frac{Q}{|\phi^2|^2} \;\, \approx \, \frac{Q}{|\phi^2|^2}  \qquad\mbox{for}\,\,\frac{Q}{|\phi^2|^2} \gg 1
\,. 
\end{align}
This is consistent with IIB supergravity \cite{Buchbinder:2001ui,Maldacena:1997re}, 
and as is well-known, reduces to $AdS^5 \times S^5$ in the near-horizon limit. 
The result above is in fact more general, since the effective 4-dimensional brane metric $G_{\mu\nu}$ is not necessarily flat. 
Therefore the quantum corrections to the matrix model can be interpreted in terms of a modified bulk geometry of the embedding space $\R^{10}$. 

Note that in contrast to previous work on related issues such as~\cite{Maldacena:1999mh,Alishahiha:1999ci,Berman:2000jw,Liu:2000ad}, the results above are exclusively based on matrix model computations, without presupposing any relation with supergravity.

More general backgrounds, which in contrast to the situation above break the $SU(N)$ gauge symmetry completely, and which can be viewed either as product space $\cM^4 \times \cK$, or in terms of a $SU(N)$ gauge theory on $\cM^4$ in the Higgs phase, may also be considered. 
Such backgrounds will typically break supersymmetry and hence lead to induced gravity terms. 
The non-Abelian fields $\cA^a=\cA^a_\a\l^\a$ may then describe a quantized compact $2n$-dimensional symplectic space $\cK$. 
Hence, the background can be interpreted as higher-dimensional space $\cM = \cM^4 \times \cK$. 
It is well-known that such $\cK$ may indeed arise from non-Abelian fields on $\cM^4$ via the Higgs effect, e.g. the fuzzy sphere $S^2_N$~\cite{Madore:1991,Myers:1999ps,Alekseev:1999,Aschieri:2006uw}. 
Some preliminary studies in this direction have been recently done in Ref.~\cite{Blaschke:2011qu}. 
In particular, the effective action on $\cM^4\times\cK_N$ in the presence of fuzzy extra dimensions was studied at different scales: 
Supersymmetry can be broken by the extra dimensions $\cK_N$ and their Kaluza-Klein modes, and 
finite gravitational terms are generically induced in the trace--$U(1)$ sector. 
Maximal supersymmetry is restored above a certain scale, ensuring a UV finite effective action.

\section{Conclusion}
We have presented a brief summary of recent results concerning the effective action of the IKKT resp. the IIB model~\cite{Ishibashi:1996xs} in the presence of dynamical SUSY breaking. 
These should illustrate (and to some extent support) the idea of a brane-world scenario with compactified extra dimensions and emergent gravity on the branes. 
Further details may be found in Refs.~\cite{Blaschke:2011qu,Chatzistavrakidis:2011gs,Steinacker:2011wb}. 
Future work on this subject will surely shed more light on this model, as there is much more to study, especially concerning compactified brane solutions of type $\cM^4\times\cK$.

\subsection*{Acknowledgements}
Many thanks go to the organizers of the 7$^{\rm th}$ International Conference on Quantum Theory and Symmetries in Prague, which was a wonderful and stimulating conference. 
D.N. Blaschke is a recipient of an APART fellowship of the Austrian Academy of Sciences.

\bibliographystyle{../../custom1.bst}
\bibliography{../../articles.bib,../../books.bib}

\end{document}